\documentclass[final,numberedheadings,letterpaper]{aipproc}

\layoutstyle{6x9}

\usepackage{graphicx,bbm,amssymb,amsopn}

\newcommand{\un}[1]{{\underline{#1}}}

\newcommand{\xbra}[1]{{( #1 \vert}}
\newcommand{\ket}[1]{{\vert #1 \rangle}}

\newcommand{\xbraket}[2]{( #1 \vert #2 \rangle}

\newcommand{\ZZ}{\mathbb{Z}}

\def\tr{{\,{\rm tr}\,}}

\begin{document}

\title{Third quantization}


\author{Thomas H. Seligman}{
 address={Instituto de Ciencias F\'isicas
	Universidad Nacional Aut\'onoma de M\'exico, Cuernavaca, M\'exico},
	altaddress={Centro Internacional de Ciencias, 
	Cuernavaca, Morelos, M\'exico}
}

\author{Toma\v z Prosen}{
  address={Department of physics, FMF, University of Ljubljana, Ljubljana, Slovenia}
}

\begin{abstract}
The basic ideas of second quantization and Fock space are extended 
to density operator states, used in treatments of open 
many-body systems. This can be done for fermions and bosons.
While the former only requires the use of a non-orthogonal basis,
the latter 
requires the introduction of a dual set of spaces. In both cases an
 operator algebra closely resembling the canonical one is developed and 
used to define the dual sets of bases. We here concentrated on 
the bosonic case where the unboundedness
of the operators requires the definitions of dual spaces to support the 
pair of bases. Some applications, mainly to non-equilibrium steady states,
will be mentioned.
  
   \end{abstract}

\maketitle


\section{Introduction}

  Second quantization has been one of the early fields of Marcos Moshinsky and his book 
on its applications to fermionic systems \cite{marcos2Q} has formed many of us. One of us 
had the pleasure to look with Marcos into the extensions of this formalism to 
non-orthogonal bases and the introduction of a dual basis and corresponding 
dual sets of operators in this context \cite{MSann}.
Since these concepts have become very useful for open many-body systems and 
recently one of us has generalized them to what may be called "third quantization" 
by applying second quantization and the Fock space concept to density operator spaces of mixed 
states and spaces of observables of fermionic systems \cite{prosen-njp}. The use of a set of dual bases 
becomes essential to maintain a simple algebraic form for the  problem.
Third quantization was successfully used to explicitly and elegantly solve situations 
with quadratic (or quasi-free) Hamiltonians and linear coupling to an environment 
via the Lindblad operators \cite{prosen-njp,prosen-prl} or via the Redfield model 
\cite{prosen-njp2}. The next step is to follow Marcos Moshinsky to the harmonic 
oscillator \cite{Marcos-ho}, i.e. to bosonic systems 
\cite{ps10}. This step implies not only the introduction of a dual basis, but actually 
of a dual space. We shall give a view of these recent developments emphasizing 
the elegance and efficiency of the methods more than any specific application. 
We shall present basis sets for the dual bases in a rather formal way and then 
show that it is the algebraic structure that determines the actual implementation.
The implication, which this has in terms of enveloping algebras, their underlying 
Lie algebras and the corresponding groups has not been explored at all. 
Furthermore the possibilities to use graded algebras and combining fermionic 
and bosonic degrees of freedom in such a treatment, is very enticing to any 
disciple of Marcos Moshinsky.  
We shall here try to lay the foundations and give a taste of things to come, 
in the hope of enticing other participants of this meeting to 
help explore this line of thinking. 
Proceeding in reverse order of temporal development, we shall first present the 
bosonic case, with its richer structure and then go to the fermionic problem. 
Finally an outlook will be given
on what should be done to obtain a  more complete theoretical descriptions of 
each of the cases and for their algebraic unification.

\section{Bosonic systems}
 
Let us consider a Hilbert-Fock space ${\cal H}$ of $n$ {\em bosons}. Elements of ${\cal H}$ 
can be generated from a particular element
$\psi_0\in{\cal H}$, called {\em a vacuum pure state}, and a set of $n$ unbounded operators over 
${\cal H}$, $a_1,\ldots,a_n$, which annihilate $\psi_0$, $a_j |\psi_0\!\!>\, = 0$, and their Hermitian adjoints
$a^\dagger_1,\ldots,a^\dagger_n$, satisfying {\em canonical commutation relations} (CCR)
\begin{equation}
[a_j,a^\dagger_k] = \delta_{j,k}, \quad [a_j,a_k] = [a^\dagger_j,a^\dagger_k] = 0.
\label{eq:CCR}
\end{equation}
Let us define a pair of vector spaces ${\cal K}$ and ${\cal K}'$, such that 
${\cal K}$ contains trace class operators, such as density matrices, and
${\cal K}'$ contains unbounded operators representing physical observables. 
We choose a specific space of observables ${\cal K'}$ and define a subspace ${\cal K}$ of 
{\em trace class} operators over ${\cal H}$, such that $\rho \in {\cal K}$ if and only if
$A \rho$ is trace class for any $A \in {\cal K}'$. Thus ${\cal K}'$  and ${\cal K}$ form a 
dual pair of Hilbert spaces and we will later choose a dual pair of bases, one from each of these spaces.

For instance, we may chose ${\cal K}'$ as a linear space of all ({\em unbounded}) 
operators whose phase space representation of the operator is an entire 
function on the corresponding $2n-$dimensional phase space.
Then ${\cal K}$ must be restricted to operators with finite support in the number 
operator basis, i.e. to operators which have a finite number of non-vanishing 
matrix elements in this basis. Such a constraint on density matrices may be too 
restrictive for certain applications. We shall show later [using eq. (\ref{eq:bases})] 
how this restriction can be relaxed by 
appropriately restricting ${\cal K'}$.
 
We now proceed with an algebraic development and conveniently adopt Dirac 
notation. We write an element of ${\cal K}$ as {\em ket} $\ket{\rho}$ and an 
element of ${\cal K}'$ as {\em bra} $\xbra{A}$, and define their contraction 
or scalar product to give the
expectation value of $A$ for a state $\rho$,
\begin{equation}
\xbraket{A}{\rho} = \tr A\rho.
\label{eq:spr}
\end{equation}
We use distinct types of brackets to emphasize the difference between the spaces from 
which the ket and the bra have to be chosen.

If $b$ is any of the operators $a_j,a^\dagger_j$, then for each $\rho \in {\cal K}$ and $A \in {\cal K}'$, 
$b \rho, \rho b$ and $A b, b A$ are also elements of ${\cal K}$ and ${\cal K}'$ respectively.
Thus we define the left multiplication maps $\hat{b}^{\rm L}$  and the right multiplication maps 
$\hat{b}^{\rm R}$ over ${\cal K}$ by
\begin{equation}
\hat{b}^{\rm L} \ket{\rho} = \ket{b \rho},\quad \hat{b}^{\rm R} \ket{\rho} = \ket{\rho b}.
\label{eq:left}
\end{equation}
The action of their adjoint on ${\cal K}'$ is defined by (\ref{eq:spr}) 
and the fact that the trace is cyclic. Thus
\begin{equation}
\xbra{A} \hat{b}^{\rm L}  = \xbra{A b},\quad \xbra{A} \hat{b}^{\rm R} = \xbra{b A}.
\label{eq:right}
\end{equation}
Loosely speaking, we can also say that ${(\hat{b}^{\rm L})}^* = \hat{b}^{\rm R}$ 
and ${(\hat{b}^{\rm R})}^* = \hat{b}^{\rm L}$.

Next we define the set of $4n$ maps $\hat{a}_{\nu,j},\hat{a}'_{\nu,j},j=1,\ldots,n,\nu=0,1$,
\begin{eqnarray}
\hat{a}_{0,j} &=& \hat{a}^{\rm L}_j ,\,\;\qquad   \hat{a}'_{0,j} =  
\hat{a^\dagger}^{\rm L}_j -  \hat{a^\dagger}^{\rm R}_j, \cr
\hat{a}_{1,j} &=& \hat{a^\dagger}^{\rm R}_j ,\qquad   \hat{a}'_{1,j} = 
 \hat{a}^{\rm R}_j -  \hat{a}^{\rm L}_j.
\label{eq:maps}
\end{eqnarray}
that have the unique properties: (i) almost-canonical commutation relations
\begin{equation}
[\hat{a}_{\nu,j},\hat{a}'_{\mu,k}] = \delta_{\nu,\mu}\delta_{j,k}, 
\quad [\hat{a}_{\nu,j},\hat{a}_{\mu,k}] =  [\hat{a}'_{\nu,j},\hat{a}'_{\mu,k}]  = 0,
\label{eq:almostCCR}
\end{equation}
(ii) $\hat{a}'_{\nu,j}$ left-annihilate the identity operator
\begin{equation}
\xbra{1}\hat{a}'_{\nu,j} = 0
\label{eq:leftvacuum}
\end{equation}
and (iii) $\hat{a}_{\nu,j}$ right-annihilate the vacuum pure state $\ket{\rho_0} 
\equiv |\psi_0\!\!> <\!\!\psi_0|$
 \begin{equation}
\hat{a}_{\nu,j}\ket{\rho_0} = 0.
\label{eq:rightvacuum}
\end{equation}
Writing a $2n$ component multi-index $\un{m}=(m_{\nu,j}\in\ZZ_+;\nu\in\{0,1\},j\in\{1\ldots n\})^T$ 
we define a dual pair of Fock bases, one for each of 
the spaces ${\cal K},{\cal K}'$ as
\begin{equation}
\ket{\un{m}} = \prod_{\nu,j} \frac{(\hat{a}'_{\nu,j})^{m_{\nu,j}}}{\sqrt{m_{\nu,j}!}} \ket{\rho_0},\quad
\xbra{\un{m}} = \xbra{1}\prod_{\nu,j} \frac{(\hat{a}_{\nu,j})^{m_{\nu,j}}}{\sqrt{m_{\nu,j}!}} 
\label{eq:bases}
\end{equation}
Their bi-orthonormality $\xbraket{\un{m}'}{\un{m}} = \delta_{\un{m}',\un{m}}$ is directly
guaranteed by the almost-CCR (\ref{eq:almostCCR}).
Here and for the rest of the paper, $\un{x}=(x_1,x_2,\ldots)^T$ designates a vector (column) 
of any, scalar-, operator- or map-valued symbols.

The explicit construction of the bases (\ref{eq:bases}) allows us to enlarge and restrict the
spaces ${\cal K}$ and ${\cal K}'$ such as to keep duality by always restricting one when 
extending the other or vice versa appropriately.
We achieve this identifying the space ${\cal K}$ with the $l^2$ Hilbert space of vectors of 
coefficients $\{ \sigma_{\un{m}} \}$, 
${\cal K} \ni \ket{\sigma} = \sum_{\un{m}} \sigma_{\un{m}} \ket{\un{m}}$, and the space 
${\cal K}'$ with the $l^2$ Hilbert space of vectors of coefficients
$\{ S_{\un{m}} \}$, ${\cal K}' \ni \xbra{S} = \sum_{\un{m}} S_{\un{m}} \xbra{\un{m}}$. 
Then, clearly by Cauchy-Schwartz inequality, 
$|\tr S\sigma| = |\sum_{\un{m}} S_{\un{m}}\sigma_{\un{m}}| < \infty$ and hence ${\cal K}$ 
and ${\cal K}'$ are dual in the required sense.

The main idea of application of the third quantization is then to express the 
generators of quantum master equations, governing the dynamics of a density matrix, 
in terms of canonical operator maps $\hat{a}_{\nu,j},\hat{a}'_{\nu,j}$. For quadratic 
systems, for example, such generators are again quadratic and can be decomposed to 
normal (master) modes, and thus diagonalized, by means of a non-unitary analogue of 
the Bogoliubov-de Gennes transformation (see \cite{prosen-njp,prosen-njp2,ps10}).
In particular, the physically interesting {\em non-equilibrium-steady-state} can be 
constructed as the {\em right-vacuum} state of our theory.

\section{Fermionic operators}

In \cite{prosen-njp} the entire idea of third quantization
was first presented, but the maps on the operator spaces were not given in terms of 
raising and lowering operators but in terms of Hermitian "coordinates" and "momenta". 
We shall here briefly indicate, that it can be done equally for fermionic (anti-commuting) 
 raising and lowering operators. Thus the situation is maintained
as symmetric as possible to the boson case. The main difference will 
naturally be, that the two bi-orthogonal bases will span the same 
Hilbert space rather than a dual pair.

Thus we point out an equivalent version of fermionic third quantization, which follows 
exactly the steps of the present communication, but starting instead from a set of 
fermionic operators $c_j, c^\dagger_j$, obeying
{\em canonical anti-commutation relations} (CAR),
\begin{equation}
\{c_j,c^\dagger_k\}= \delta_{j,k}, \quad \{c_j,c_k\} = \{c^\dagger_j,c^\dagger_k\} = 0,
\label{eq:CAR}
\end{equation}
introducing the dual sets of density operators and observables, stating (\ref{eq:spr},
\ref{eq:left},\ref{eq:right}) and defining the canonical adjoint fermionic maps
\begin{eqnarray}
\hat{c}_{0,j} &=& \hat{c}^{\rm L}_j ,\;\quad\qquad   \hat{c}'_{0,j} =  
\hat{c^\dagger}^{\rm L}_j -  \hat{c^\dagger}^{\rm R}_j \hat{\cal P}, \cr
\hat{c}_{1,j} &=& \hat{c^\dagger}^{\rm R}_j \hat{\cal P} ,\qquad   \hat{c}'_{1,j} =  
\hat{c}^{\rm R}_j \hat{\cal P} -  \hat{c}^{\rm L}_j,
\label{eq:fmaps}
\end{eqnarray}
satisfying almost-CAR
\begin{equation}
\{\hat{c}_{\nu,j},\hat{c}'_{\mu,k}\} = \delta_{\nu,\mu}\delta_{j,k}, 
\quad \{\hat{c}_{\nu,j},\hat{c}_{\mu,k}\} =  \{\hat{c}'_{\nu,j},\hat{c}'_{\mu,k}\} = 0,
\label{eq:almostCAR}
\end{equation}
and the properties (\ref{eq:leftvacuum},\ref{eq:rightvacuum}). The parity superoperator 
$\hat{\cal P}$ is uniquely defined by its action on the dual vacuum states, 
$\xbra{1}\hat{\cal P} = \xbra{1}, 
\hat{\cal P}\ket{\rho_0} = \ket{\rho_0}$, and requiring that it anti commutes with all the elements of the 
adjoint-algebra $\{\hat{\cal P},\un{\hat c}\} = \{\hat{\cal P},\un{\hat c}'\} = 0$.
The difference to the more symmetric approach \cite{prosen-njp} is 
that now the canonical conjugate 
adjoint maps are {\em not} the hermitian adjoint maps
$\hat{c}'_{\nu,j} \neq \hat{c}^\dagger_{\nu,j}$, which is however 
of no consequence as we are 
anyway dealing with problems in which
{\em non-normal} operators enter in an essential way.
The complete bi-orthogonal bases of the density operator space ${\cal K}$, 
and its dual, the 
space of observables ${\cal K}'$, analogue to (\ref{eq:bases}), 
can now be labelled by means
of binary multi indices $\un{m}$, $m_{\nu,j}\in\{0,1,\}$,
\begin{equation}
\ket{\un{m}} = \prod_{\nu,j} (\hat{c}'_{\nu,j})^{m_{\nu,j}} \ket{\rho_0},\quad
\xbra{\un{m}} = \xbra{1}\prod_{\nu,j} (\hat{c}_{\nu,j})^{m_{\nu,j}}
\label{eq:bases1}
\end{equation}

\section {conclusions}
We have presented third quantization for bosonic and fermionic operators
states, attempting a uniform presentation. The result in itself is remarkable and 
very useful, but was essentially taken form refs \cite{prosen-njp,prosen-njp2,ps10}. 
The purpose of this presentation was to show the structure of the formalism
presented in a form which prepares the use of algebraic and group theoretical 
techniques along the lines developed by Marcos Moshinsky. Operator states were 
not known and non-orthogonal bases were rarely used in this context, but in 
\cite{MSann} Marcos Moshinsky together with one of the authors developed 
the latter for fermionic systems in second quantization. Even there advantage was taken of the fact, that the 
algebraic structure defines the important features of the problem.
Thus combining raising operators in the dual basis with lowering operators in the 
original one, one could mimic the algebraic structure developed for
standard anti commuting operators. This clearly carries over to the third quantized 
picture. For bosonic systems the situation is a little more involved because we deal 
with operators on different spaces, yet we feel confident, that we can develop the 
techniques  known as dynamical algebras or spectrum generating algebras also in this 
case, though the central interest clearly is not on any spectrum. The entire idea 
relies on the point that through bases  (\ref{eq:bases}) and (\ref{eq:bases1}) we can 
pass to equivalent infinite  $l^2$ spaces in the first case and finite ones in the second. 
Limiting algebraic operations appropriately we never see the differences or they
appear in the occasional use of an overlap matrix or its inverse.
Note that the freedom we have in the fermionic case is larger than in the bosonic one, 
as we are never in danger of leaving the finite dimensional Hilbert space.

Summarizing we may say that the construction we present provides an ideal framework for 
algebraic or group-theoretical developments. Yet filling the framework has barely begun.                                                                                                                                                                                                                                     
It also seems obvious that we can construct graded algebras also known as superalgebras, 
mixing anti-commuting and commuting variables, in the present context.


\begin{theacknowledgments}

  We acknowledge discussions with F. Leyvraz and J. Eisert.
This work was supported by the Programme P1-0044, and the Grant J1-2208, of Slovenian Research Agency, 
and by CONACyT, Mexico, project 57334 as well as the University of Mexico, PAPIIT project 
IN114310.

\end{theacknowledgments}

\bibliographystyle{aipproc}

\end{document}